# Accelerator performance analysis of the Fermilab Muon Campus

Diktys Stratakis, Mary E. Convery, Carol Johnstone,
John Johnstone, James P. Morgan, and Dean Still
*Fermi National Accelerator Laboratory, Batavia, Illinois 60510, USA*

Jason D. Crnkovic, Vladimir Tishchenko, and William M. Morse
*Brookhaven National Laboratory, Upton, New York 11973, USA*

Michael J. Syphers
*Northern Illinois University, DeKalb, Illinois 60115, USA,
and Fermi National Accelerator Laboratory, Batavia, Illinois 60510, USA*

Fermilab is dedicated to hosting world-class experiments in search of new physics that will operate in the coming years. The Muon g-2 Experiment is one such experiment that will determine with unprecedented precision the muon anomalous magnetic moment, which offers an important test of the Standard Model. We describe in this study the accelerator facility that will deliver a muon beam to this experiment. We first present the lattice design that allows for efficient capture, transport, and delivery of polarized muon beams. We then numerically examine its performance by simulating pion production in the target, muon collection by the downstream beam line optics, as well as transport of muon polarization. We finally establish the conditions required for the safe removal of unwanted secondary particles that minimizes contamination of the final beam.

## I. INTRODUCTION

Measurement of the muon anomalous magnetic moment, $\alpha_\mu = (g_\mu - 2)/2$, provides an important test of the Standard Model (SM) [1]. The Brookhaven Muon g-2 Experiment measured a discrepancy of more than 3 standard deviations compared to the SM prediction [2]. As this discrepancy hints at the possibility of new physics, efforts have been made to improve both the theoretical [3] and experimental [4,5] precision of $\alpha_\mu$.

The Fermilab Muon g-2 Experiment [6] is a next generation experiment mainly motivated by this discrepancy. The goal of the experiment is a fourfold improvement in the measured precision that will reduce the experimental uncertainty to 140 parts per billion (ppb) [7]. One of the key requirements to achieve this goal is a factor of 20 increase in recorded muon decays compared to the Brookhaven experiment [8].

A combination of beam lines that are part of the so-called Muon Campus [9] have been designed to deliver a muon beam with sufficient intensity. The sequence starts at the booster [10], where short batches of 8 GeV protons are delivered to the recycler ring [11]. An rf system separates a batch into four tighter bunches of $10^{12}$ protons each. These bunches are extracted one at a time and directed to a target station, which is tuned to collect 3.1 GeV/$c$ positive secondary particles. Pions and their daughter muons are transported along a 280 m channel and then injected into the repurposed antiproton debuncher ring [12,13], where they make several revolutions before being kicked into a final beam line that terminates at the entrance of the muon storage ring that is used by the Muon g-2 Experiment.

The statistical uncertainty of the measurement scales as $1/\sqrt{N}P^2$ [14], where $P$ is the beam polarization and $N$ is the number of muons, and the muon storage ring has a narrow momentum acceptance of less than $\pm 0.5\%$. It is therefore essential to maximize polarized muon transmission along the Muon Campus, while minimizing the growth of the beam emittance. Moreover, the delivery of a clean muon beam that has a pion contaminant fraction below $10^{-5}$, with no protons present, is another key requirement, as these hadrons can cause a large hadronic "flash" at injection that paralyzes detector systems and leads to baseline shifts on a slowly decaying background. The present accelerator scheme, however, has a series of bending sections, elevation changes as well as complicated injection and extraction regions. Such beam line sections, if not well controlled, can degrade the beam quality substantially [15]. The three interesting questions then are, what is the overall transmission and momentum spread of positive muons, what is the level of contamination from protons and pions, and what is the degree of polarization of the delivered beam. A key requirement to resolve these issues is the ability to accurately evaluate the performance of the Muon Campus beam lines.

The main goal of this paper is to present the baseline lattice for the Muon Campus and evaluate its performance numerically. In order to accomplish this, we follow a sequence of steps. First, we present a lattice design to efficiently capture, transport and deliver a muon beam from the production target towards the storage ring of the Muon g-2 Experiment. Then, we numerically examine its performance using G4BEAMLINE [16], a GEANT4 [17] based code that fully incorporates all basic physical processes such as muon decay and spin precession. We initiate our tracking at the muon production target and present a detailed end-to-end simulation. In order to support our findings, we compare our numerical results against an exponential-decay estimate and find good agreement. Our study suggests that the Muon Campus has the potential to deliver highly polarized muons at a rate of $\sim 8 \times 10^5$ $\mu^+$ per bunch.

The outline of this paper is as follows: In Sec. II, we give an overview of the Muon Campus and provide details of the design parameters of all beam lines. In Sec. III we report the results from our simulations modeling the aforementioned channel. Finally, we present our conclusions in Sec. IV.

## II. LATTICE DESIGN AND OPTICS

To enhance cost savings, the existing tunnel enclosures and beam lines from the Antiproton Source [13] were largely reused for operation of the Muon Campus. In addition, stochastic cooling components and other infrastructure no longer needed in the debuncher ring were removed in order to improve its aperture and the ring was renamed delivery ring (DR).

Figure 1 displays a schematic layout of the Fermilab Muon Campus, where 8 GeV protons are transported via the M1-line to an Inconel target at AP0 for the Muon g-2 Experiment. A secondary beam from the target is collected using a lithium lens, where the positively charged particles with 3.1 GeV/$c$ ($\pm 10\%$) are selected by using a bending magnet. The secondary beam leaves the target station and travels through the M2- and M3-line, which is designed to capture as many 3.1 GeV/$c$ muons as possible from the pion decays. The beam is then injected into the DR, where essentially all remaining pions decay into muons after several revolutions. The DR is also used to separate muons in time from the heavier protons. A kicker is then used to remove the protons, where the muon beam is extracted into the M4-line. The beam then continues along the M5-line which terminates just upstream of the entrance to the muon storage ring, where the storage ring entrance is taken as the simulation end point for this study.

Figure 2(a) shows a schematic layout of the target station, which consists of five main devices: pion production target, lithium lens, collimator, pulsed magnet, and beam dump. The primary proton beam interacts with the target to produce positive secondaries that are focused by the lithium lens and then are momentum selected via the downstream pulsed dipole magnet (PMAG). The target [19] charged particles, and protons that do not interact with the target. Table I provides parameters for the beam incident on the target.

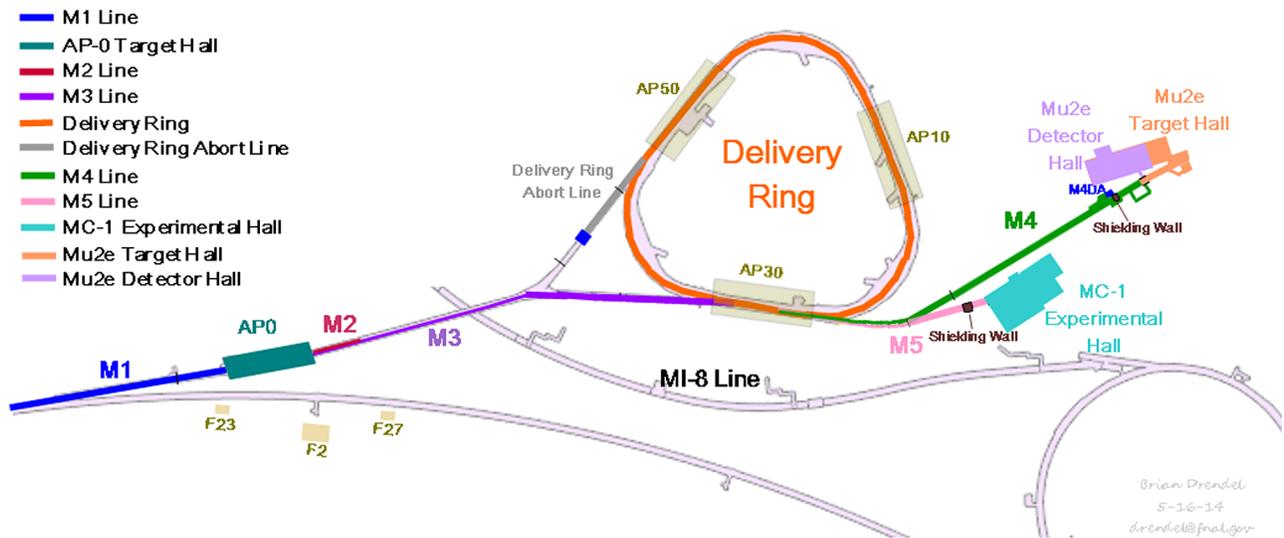

FIG. 1. A schematic representation of the Muon Campus accelerator complex that is used by the Muon g-2 Experiment. Secondaries are produced on a target that then travel through the M2- and M3-line, which is designed to capture as many 3.1 GeV/$c$ muons from pion decays as possible. The beam is injected into the DR wherein a kicker is used to remove the protons, the resulting muon beam is then extracted into the M4-line, and the muon beam is eventually transferred to the new M5-line that leads to the muon storage ring (enclosed in the MC1 hall). The M3-line, DR, and M4-line are also designed to be used for 8 GeV proton transport by the Mu2e Experiment [18]. The combined M2- and M3-line and M4-and M5-line lengths are 280 and 130 m respectively, along with the DR that has a circumference of 505 m. One should note that the horizontal alignment of the M3-line with the injection leg of the DR is achieved by two 9.25° bends and an additional 5° bend that is placed upstream of the AP30 straight section.

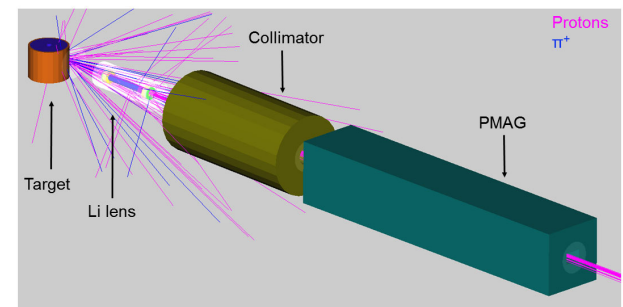

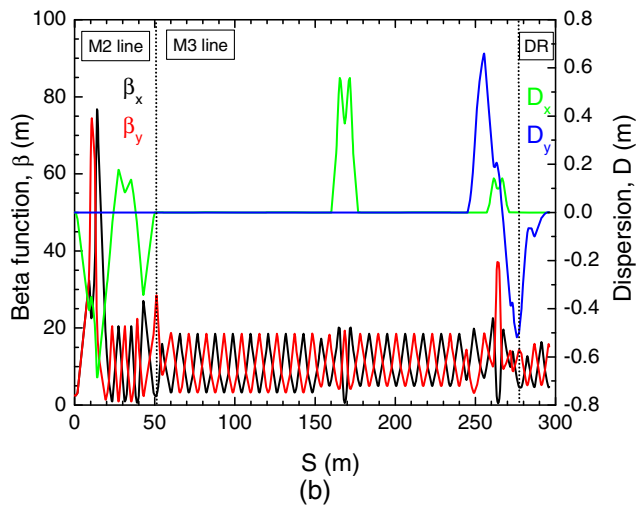

FIG. 2. (a) A Schematic representation of the Muon Campus target station, where the positive pion ($\pi^+$) and proton trajectories are shown in blue and pink respectively. (b) The optics functions are shown for the M2- and M3-line, where the lattice is designed to have a transverse acceptance of 40 mmmrad, and a narrow momentum acceptance of $\Delta p/p = \pm 2\%$. The beam line has 120° phase-advance cells between $S = 23.2$ m and 35.6 m, 90° phase-advance cells between 65.4 and 143.1 m, and 72° phase-advance cells between 177.8 to 240.2 m. The point of intersection between M2 and M3 is at $S = 50.0$ m, the two 9.25° horizontal bending magnets are at $S = 160.0$ m and $S = 174.0$ m, while injection to the DR occurs at $S = 280.0$ m.

TABLE I. Primary beam parameters.

| Parameter | Value |
|---|---|
| Intensity per pulse | $10^{12}$ |
| Total POT per cycle | $16 \times 10^{12}$ |
| Number of pulses per cycle | 16 |
| Cycle length (s) | 1.4 s |
| Primary momentum (GeV/$c$) | 8.89 |
| Beam size at target (mm) | 0.15 |

The M2-line immediately follows the target station, where it starts with four matching quadrupoles that are followed by a sequence of 120° phase-advance FODO cells that have a 3.98 m half-cell length. Figure 2(b) shows the optics functions along the M2- and M3-line, where these functions are calculated with the beam optics program MAD-X [20]. One should note that $S = 0$ is set 0.722 m upstream of the PMAG, while the M2-line and M3-line intersect at $S \approx 50.0$ m. A large aperture dipole magnet provides the second 3° bend to align the beam with the M3-line trajectory at this intersection point. There is a 540° horizontal phase advance between the PMAG dipole and the second 3° bend at the M2-M3 intersection to cancel the horizontal dispersion created by the PMAG. The M3-line continues with a sequence of 90° phase-advance FODO cells that have a 5.50 m half-cell length. The beam line is aligned with the injection leg of the DR with a horizontal bend to the right at $S = 160.0$ m, which is provided by a specialized insertion created from two 9.25° dipole bends. This line continues with another sequence of 72° phase-advance FODO cells that have a 5.61 m half-cell length. The M3 line ends at $S = 280.0$ m, wherein the beam is injected vertically into the DR with a bend produced by a combination of a C-magnet (CMAG), followed by a large-aperture focusing quad (D3Q) and a pulsed magnetic septum dipole. Figure 3(a) shows a conceptual layout of the injection scheme from the M3-line to the DR. Two kicker modules (IKIK1, IKIK2) downstream of quad D30Q close the trajectory onto the DR orbit. Figure 2(b) shows that the M2- and M3-line beta functions vary smoothly (except in the matching region near the target) helping to maximize capture of the muon beam.

The primary purpose of the 505 m circumference DR is to provide sufficient time for most pions to decay into muons. The original debuncher [12,13] lattice design is used with just a few modifications [6], where Fig. 3(c) displays a schematic layout wherein the injection and extraction beam lines are omitted for simplicity. A salient feature of the DR is that it has a periodicity of 3, where the basic arrangement of the ring includes three long dispersion-free straight sections together with three arc sections having mirror symmetry in each of the three has a cylindrical core made of a nickel-iron alloy called Inconel-600 that is surrounded by a thick casing of beryllium. The magnetic focusing lithium lens element consists of a lithium rod having a 0.01 m radius and 0.16 m length that is surrounded by a titanium casing that uses a water-cooling jacket and is capped by two beryllium windows. This element is located 0.3 m downstream of the target, and it conducts a 116 kA current that produces a 232 T/m gradient within the lithium rod.

The PMAG selects 3.1 GeV/$c$ positive particles and bends them 3° into the M2-line channel, where the PMAG is protected from excess energy deposition by a water-cooled collimator that absorbs particles. This magnet is 1.03 m long, operates with a 0.53 T field, and is placed 1.3 m from the target. A beam dump is also present, but not shown, that absorbs off-momentum particles, negatively sectors. The magnet structure consists of 57 approximately 60° phase-advance FODO cells, where the beam line's natural chromaticity is compensated by two families of sextupoles bracketing all arc quads. These arcs have a dispersion function maximum of 2.4 m and transverse beta function maxima of approximately 15 m, along with the typical tunes of 9.79 horizontal and 9.77 vertical. Figure 4(a) presents the optical functions along the DR that are calculated with MAD-X.

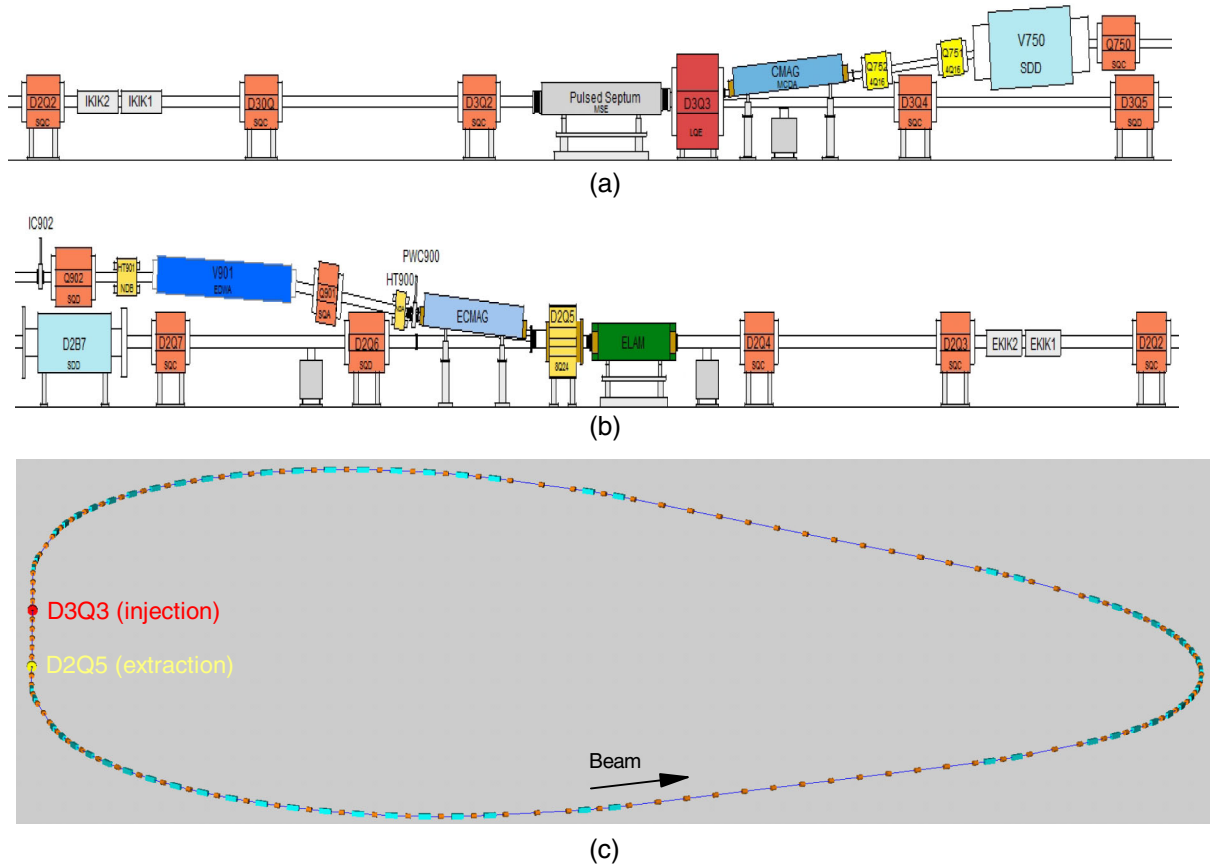

FIG. 3. (a) The conceptual design for injection into the DR, where the beam direction is from right to left. (b) The conceptual design for extraction from the DR, where the beam direction is from right to left. (c) A distorted image of the DR that shows all focusing elements, where the large aperture injection quad (D3Q3) is shown in red, extraction quad (D2Q5) is shown in yellow, all other quads are shown in orange, and bends are shown in cyan. Injection and extraction beam lines are omitted for the sake of simplicity.

Injection from the M3-line and extraction to the M4-line takes place in the same straight section with the latter happening in the downstream half, where Fig. 3(b) shows a schematic layout of the extraction scheme [6]. Two kicker magnets (EKIK1, EKIK2) upstream of quad D2Q3 are first used to kick the beam off of the closed orbit, and then a combination of three vertical bending magnets (ELAM, ECMAG, EDWA), in conjunction with an intervening quadrupole (D2Q5), bend the beam upward to a final elevation of 0.81 m above the DR. The beam bends upward into the M5-line after traversing 30 m in the M4-line, and continues towards the muon storage ring. The section of the M4-line between the DR and the beginning of the M5-line must also be able to transport 8 GeV protons, as it is shared with the Mu2e Experiment [18]. Figure 4(b) shows the beta and dispersion functions along the M4- and M5-line as calculated by MAD-X, where $S = 0$ corresponds to the front edge of the ECMAG magnet and the beam line terminates 0.30 m upstream of the entrance to the muon storage ring. The M5-line is 100 m long and includes a 27.1° horizontal bend string at $S = 46.5$ m that provides the proper entry position and angle into the muon storage ring, and right before the end of the M5-line there is a strong-focusing and tunable final focus section using four quadrupole magnets at $S = 119.8$ m that provides optical matching to the storage ring. The FODO section between the horizontal bend string and final focus section is designed to transport beam with minimal losses.

### III. TRACKING STUDIES

The performance of the Muon Campus beam lines is simulated using G4BEAMLINE, a scripting tool for the GEANT4 Monte Carlo program. The majority of the Muon Campus quadrupoles have special vacuum chambers that conform to the poles [21] in order to extend the aperture in the horizontal and vertical planes, where these apertures have been included in our G4BEAMLINE simulation model. The beam-target interactions are modeled with MARS [22] by assuming $10^9$ incident protons on the target (POT), where a particle distribution is then created 0.772 m upstream of the PMAG and propagated using G4BEAMLINE. The input distribution in the G4BEAMLINE simulation contains $1.25 \times 10^{-2}$ protons and $1.17 \times 10^{-3}$ positive pions per POT. Note that any muons produced within the lithium lens volume are excluded from the downstream simulation, since they contribute less than 3% to the total delivered muons at the entrance of the muon storage ring used by the Muon g-2 Experiment. Table II summarizes the number of secondary particles at different locations along the Muon Campus beam lines.

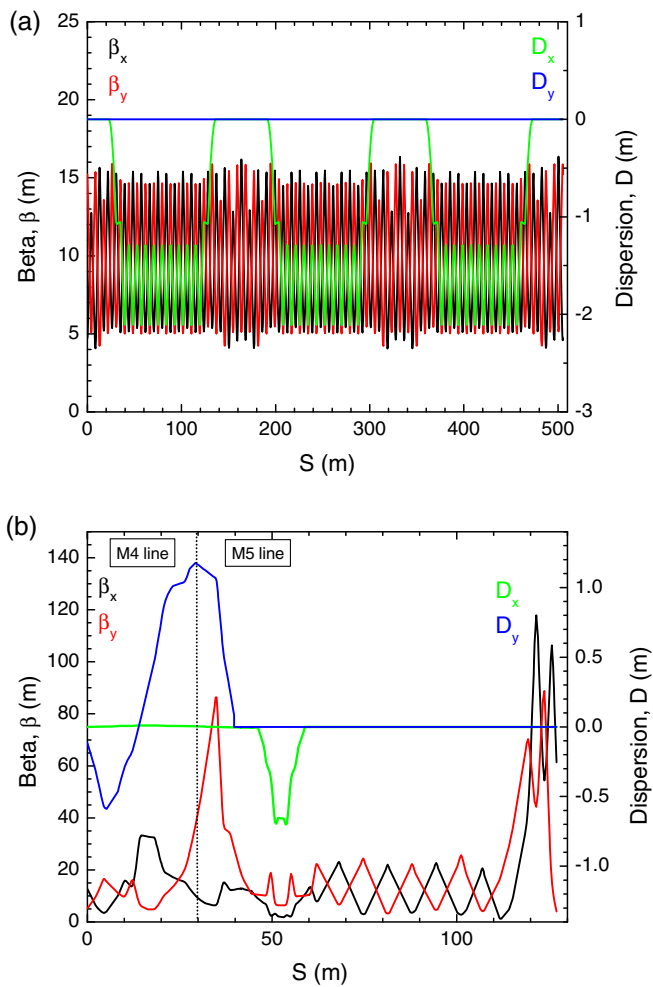

FIG. 4. (a) The optics functions for the DR. (b) The optics functions for the M4-and M5-line, where the M5-line ends at the entrance of the muon storage ring. The lattice is designed to have a transverse acceptance of 40 mm-mrad and a narrow momentum acceptance of $\Delta p/p = \pm 2\%$. The M5 begins at $S \approx 30.0$ m, a 27.1° horizontal bend string begins at $S = 46.5$ m that provides the proper entry position into the muon storage ring, while a tunable final focus section using four quadrupole magnets begins at $S = 119.8$ m that provides optical matching to the storage ring.

The evolution of secondary particles as a function of distance along the M2- and M3-line is shown in Fig. 5, where the number of muons and pions are shown with the momentum ranges of 3.1 GeV/c $\pm 2\%$ and $\pm 0.5\%$. A glance of Fig. 5 reveals a rapid loss of secondary particles within the first 10 m of the channel. This fact is not surprising since the particles produced at the target have a very wide momentum spread which extends all the way to 7 GeV/c. On the other hand, PMAG selects particles only near 3.1 GeV/c. As a result, a significant number of particles are lost from the momentum selection process. Note further that the population of muons drops roughly by a factor of 2 near $S = 8.7$ m. This is also the location of the first quadrupole magnet of M2, suggesting that collimation is an additional factor of particle loss in the first few meters of the beam line.

Further examination of Fig. 5 indicates a substantial beam loss at $S = 160.0$ m, which corresponds to the location of one of the two 9.25° horizontal bending magnets in the M3-line. While for $\Delta p/p = \pm 2\%$ and $\Delta p/p = \pm 0.5\%$ there is no pion loss at $S = 160.0$ m, the daughter muon loss is nearly 40% for both momentum acceptances suggesting that the performance deterioration for muons is primarily caused by the transverse aperture rather the momentum cut. We estimate the effective transverse beam emittance, from the second-order moments, upstream of the horizontal bend as 38 mm-mrad for pions and 120 mm-mrad for muons. As the Muon Campus lines have a 40 mm-mrad transverse acceptance [23], we conclude that the nearly 40% muon loss is from collimation due to the muon beam emittance being 3 times larger than the lattice acceptance.

The simulation finds that the beam is dominated by secondary protons at the end of the M3-line ($S_{\text{CMAG}} \approx 280.0$ m), where the distribution of muons arriving at the CMAG has an average momentum of 3.1 GeV/c with one standard deviation of 2.1%. The number of pions surpasses the number of muons by an order of magnitude, which indicates the potential for collecting more muons further downstream. There are 93,960 positive pions that survive when tracking $1.17 \times 10^6$ pions from the end of the lithium lens to the end of the M3-line and turning off pion decay (dashed line), while there are 18,698 positive pions and 2680 positive muons that survive when turning on pion decay. The triangle in Fig. 5 corresponds to the exponential decay expression $N = N_o e^{-t_{\text{CMAG}}/\gamma \tau_\pi}$ that agrees very well with our simulations, where $\tau_\pi = 26$ ns is the average rest frame pion lifetime, $\gamma$ is the Lorentz factor, $N_0$ is the number of pions without decays, and $t_{\text{CMAG}} = 933$ ns is the time required for the beam to reach the CMAG.

Figure 6 shows the number of secondary particles as a function of the number of turns in the DR. Protons outnumber both muons and pions by at least 2 orders of magnitude, which is unsurprising as a considerable number of secondary protons travel downstream of the target and the pions decay into muons, where only the daughter muons near 3.1 GeV/c are accepted. The pion population decreases substantially, and eventually drops below $10^{-9}$ per POT after three turns. Conversely, there is only a 3% muon loss after each turn, which is close to the expected 2.6% muon decay loss after each turn for the 3.1 GeV/c muons that have a 64 $\mu$s lab frame lifetime. Finally, the rms transverse beam emittance remains practically unchanged with 14.1 mm. mrad horizontal and 15.1 mm. mrad vertical values.

TABLE II. Number of secondary particles at different Fermilab Muon Campus locations, where these values are normalized to the POT.

|  | p, all | $\pi^+$, all | $\mu^+$, all | $\mu^+$, $\Delta p/p = \pm 2\%$ | $\mu^+$, $\Delta p/p = \pm 0.5\%$ |
|---|---|---|---|---|---|
| End of M3 | $1.37 \times 10^{-4}$ | $1.87 \times 10^{-5}$ | $2.68 \times 10^{-6}$ | $1.19 \times 10^{-6}$ | $3.26 \times 10^{-7}$ |
| DR (Turn 1) | $8.02 \times 10^{-5}$ | $5.06 \times 10^{-7}$ | $9.50 \times 10^{-7}$ | $8.40 \times 10^{-7}$ | $2.59 \times 10^{-7}$ |
| DR (Turn 2) | $7.94 \times 10^{-5}$ | $2.72 \times 10^{-8}$ | $9.15 \times 10^{-7}$ | $8.16 \times 10^{-7}$ | $2.54 \times 10^{-7}$ |
| DR (Turn 3) | $7.89 \times 10^{-5}$ | $1.93 \times 10^{-9}$ | $8.83 \times 10^{-7}$ | $7.89 \times 10^{-7}$ | $2.47 \times 10^{-7}$ |
| DR (Turn 4) | $7.88 \times 10^{-5}$ | $<10^{-9}$ | $8.54 \times 10^{-7}$ | $7.65 \times 10^{-7}$ | $2.39 \times 10^{-7}$ |
| End of M5 |  | $<10^{-9}$ | $7.80 \times 10^{-7}$ | $6.80 \times 10^{-7}$ | $2.08 \times 10^{-7}$ |

Figure 7 gives the muon transmission along the M4- and M5-line after four revolutions in the DR for several different momentum cuts, where in a fashion similar to Fig. 4(b), the upstream face of the ECMAG is at $S = 0$ and the simulation terminates at the end of the M5-line. We found that > 85% of the muons reach the end of the line, where the primary sources of muon loss are collimation in the vertical extraction section and horizontal bend along the

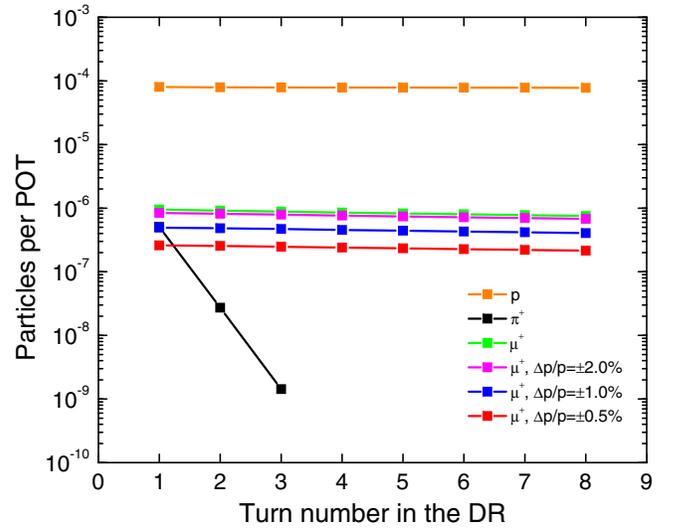

FIG. 6. The simulated performance along the DR, where the evolution of secondary particles is given as a function of the number of turns for different momentum acceptances. One should note that all pions decay to muons after the third turn.

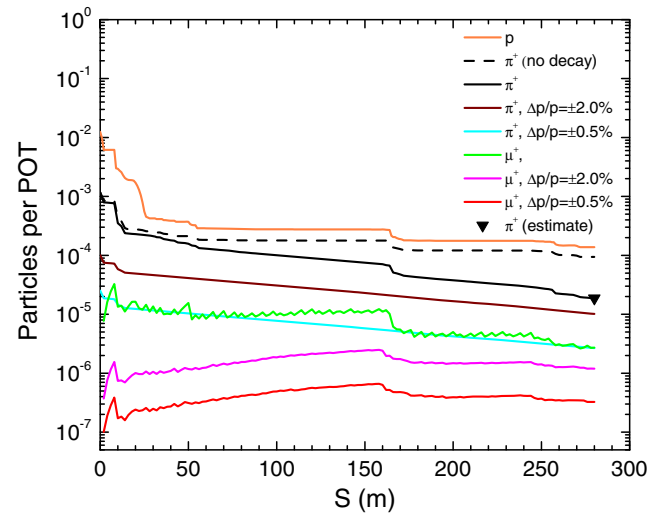

FIG. 5. The simulated performance along the M2- and M3-line, where the evolution of secondary particles is shown as a function of distance along the channel for different momentum acceptances. The dashed line shows the performance of pions when decays are not included. The final number of pions is $1.87 \times 10^{-5}$ per POT at the end of the channel ($S = 280$ m), which is in agreement with our theoretical findings (triangle). One should note that $S = 0$ is the same point as in Fig. 2(b).

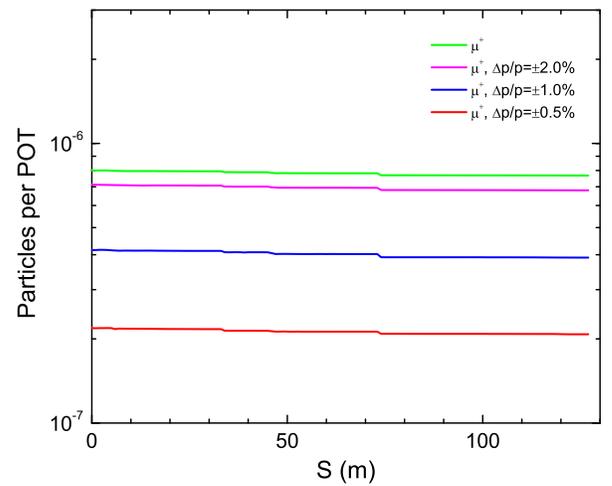

FIG. 7. The simulated performance along the M4-and M5-line, where the evolution of muons is given as a function of distance along the channel for different momentum acceptances.

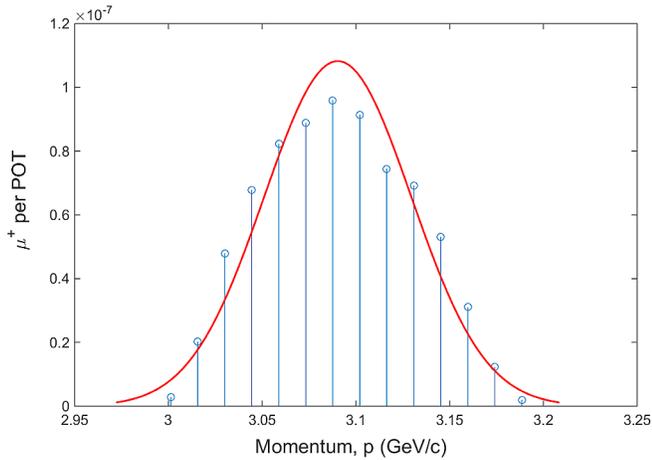

FIG. 8. The muon momentum distribution at the end of the Muon Campus, where the muons have an average momentum of 3.09 GeV/c with one standard deviation of 1.20%. The solid curve is a Gaussian distribution fit to the points.

M4- and M5-line. The total number of muons at the end of the M5-line is $\sim 8 \times 10^{-7}$ per POT or $8 \times 10^5$ per proton bunch on target (per fill), while the horizontal and vertical rms beam emittances are 14.0 mm-mrad and 15.0 mm-mrad respectively. Our simulation also successfully captures the muon momentum distribution spread, where Fig. 8 shows that the muon distribution at the end of the M5-line has an average momentum of 3.09 GeV/c with one standard deviation of 1.20%. Based on the published parameters in Refs. [6,24], $1.095 \times 10^{14}$ muons within $\Delta p/p = \pm 2.0\%$ at the end of the M5-line are required for the collection of the required Muon g-2 Experiment statistics [25,26]. From the simulation results of Fig. 7, the number of muons within $\Delta p/p = \pm 2\%$ is $6.8 \times 10^{-7}$ per POT or $6.8 \times 10^5$ per fill. Given that 16 fills occur in 1.4 s (see Table I), the Muon Campus has the potential to deliver the desired muons within about a year of running time. Moreover, the performance of the Muon Campus beam lines was modeled [27] with BMAD, an independent simulation code that also included muon decay. Both BMAD and G4BEAMLINE revealed good agreement (within 5%) for the predicted number of muons at the end of M5.

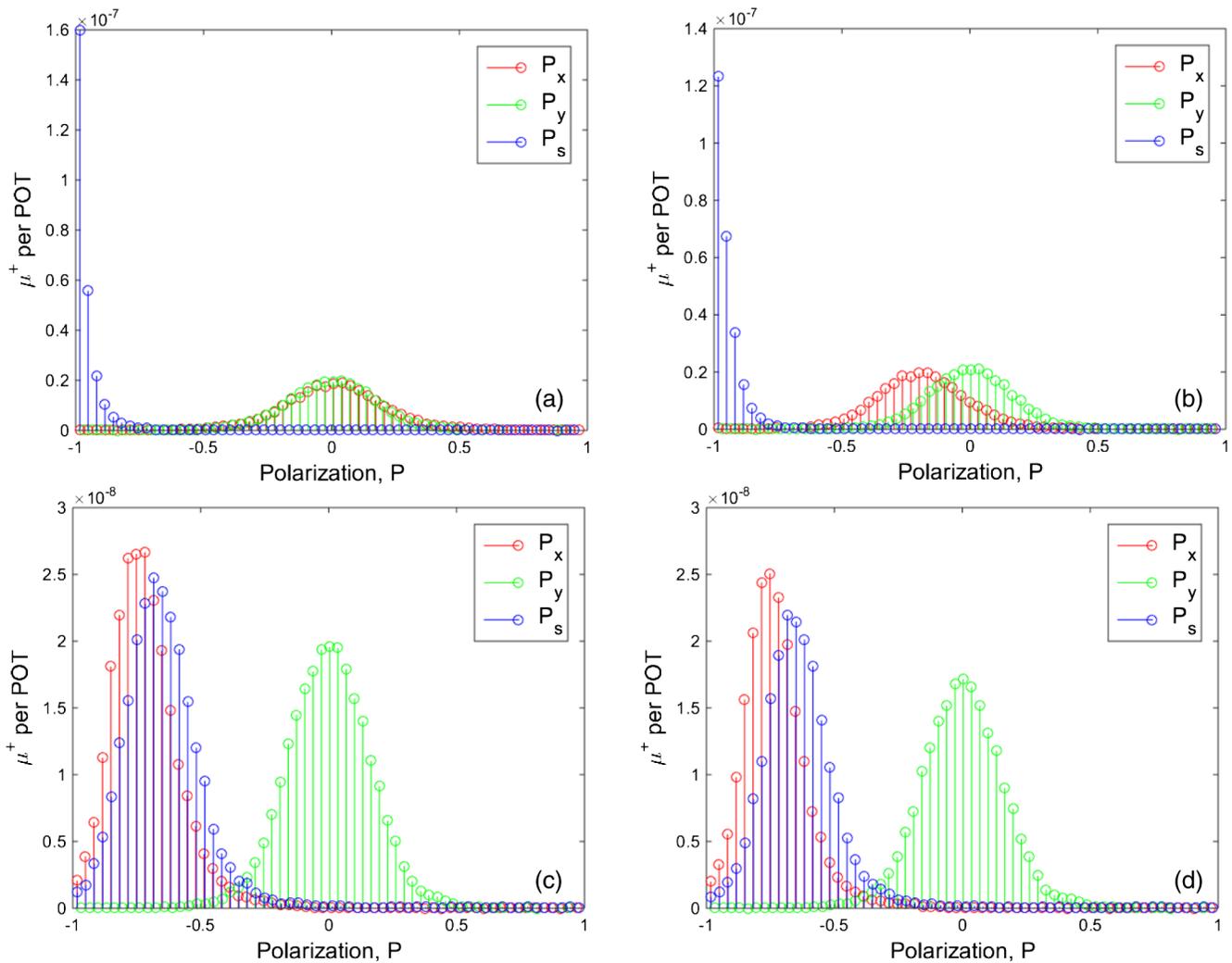

FIG. 9. The simulation results for the muon horizontal ($P_x$), vertical ($P_y$), and longitudinal ($P_S$) polarizations: (a) at the end of M3-line, (b) after one turn in the DR, (c) after four turns in the DR, and (d) at end of the M5-line. The average muon beam polarization is ~96% and remains unchanged through the channel.

TABLE III. Average muon beam horizontal and longitudinal polarization at different Muon Campus locations.

| Location | $\langle P_x \rangle$ | $\langle P_s \rangle$ |
|---|---|---|
| End of M3 | −0.01 | −0.96 |
| DR (Turn 2) | −0.38 | −0.88 |
| DR (Turn 4) | −0.71 | −0.65 |
| End of M5 | −0.72 | −0.64 |

Polarized muons are obtained from the weak decays of in flight pions: $\pi^+ \rightarrow \mu^+ + \nu_\mu$. The daughter muons have a very wide momentum spectrum in the lab frame that ranges from around one-half of the pion momentum (backward decays) to slightly greater than the pion momentum (forward decays), where forward and backward refers to the center-of-mass frame muon direction relative to the Lorentz boost between the frames. Figure 9 in combination with Table III show the muon polarization evolution along different locations within the Muon Campus. Numerical simulation predicts that 70% of the pions have decayed into muons when the beam reaches the end of the M3-line, where only forward muon decays are eventually selected given the narrow $\Delta p/p = \pm 2\%$ momentum acceptance of the channel. The muon beam has an average longitudinal polarization of 96% when it reaches the DR [Fig. 9(a)], where the polarization is negative due to the positive muon spin and momentum having opposite directions from parity violation. We find good agreement [28] when comparing our simulation results to the theoretical expressions given by Combley and Picasso [29]. The beam experiences a vertical magnetic field when it circulates in the DR, which causes the muon spins to precess in the horizontal plane [Fig. 9(b)]. The average polarization is almost equal in both horizontal and longitudinal directions after four revolutions [Fig. 9(c)], and it remains practically unchanged as the beam travels through the M4- and M5-line [Fig. 9(d)]. The final results align well with the desired parameters in Ref. [6], which specifies an average polarization greater than or equal to 90% for the Muon g-2 Experiment.

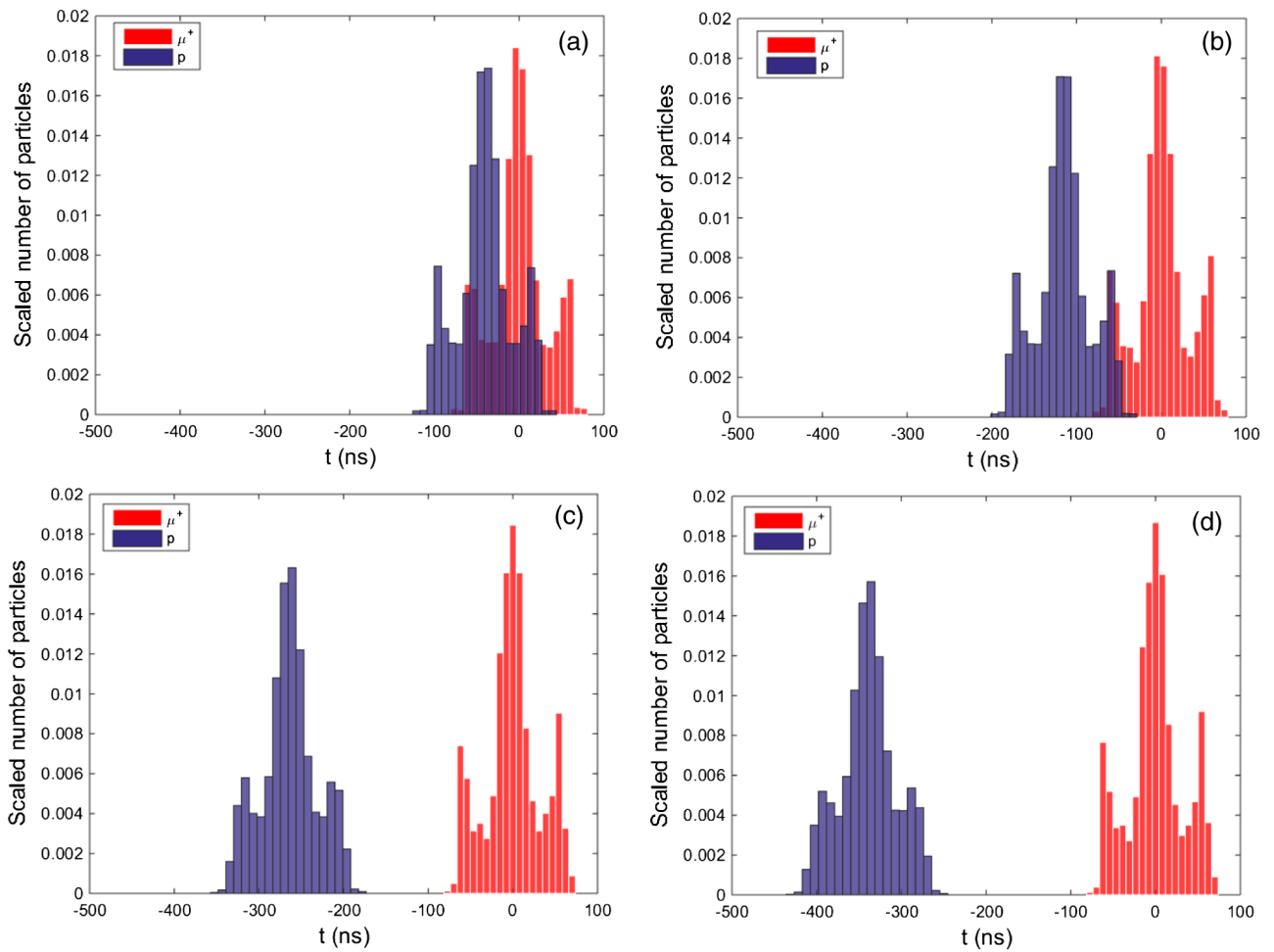

FIG. 10. The tracking of muon and proton longitudinal distributions along the DR: (a) entering, (b) after one turn in, (c) after three turns in, and (d) after four turns in the DR. The separation of muon and proton bunches increases by 75 ns after each turn. This plot indicates that four revolutions are required to safely remove the proton contamination, because of the ∼180 ns rise time for the proton removal kicker magnets. One should note that the histograms are normalized so that the area sum is equal to 1.

A complete removal of protons before injection into the muon storage ring is required in order to reduce contamination, which is accomplished by means of kicker magnets [6,23] that are strategically located in one of the DR straight sections. The proton removal kicker magnet rise times are ∼180 ns [6,30], and must be taken into account to achieve a clean separation between muons and protons. We use G4BEAMLINE to determine the appropriate timing by tracking the muon and proton populations around the DR and by recording the longitudinal distributions as a function of the number of turns. The proton and muon bunches both retain the length distribution of the incoming proton beam from the recycler [31], where the two types of bunches almost overlap as they enter the DR [Fig. 10(a)]. The two types of bunches begin to separate only after the first turn [Fig. 10(b)], due to the muons having a higher velocity that arises from having a lower mass. The nominal DR revolution times are 1.685 and 1.760 $\mu$s for 3.1 GeV/$c$ muons and protons respectively, whereby the gap size between the two bunches is expected to increase by 75 ns per turn, which is consistent with our findings. As gap size, we define the distance between the left end of the muon bunch and the right front of the proton bunch. The gap size is ∼135 ns after three turns [Fig. 10(c)], and grows to ∼210 ns after four turns [Fig. 10(d)]. These simulation results imply that at least four turns in the DR are needed in order to fully remove protons without any significant muon losses.

## IV. SUMMARY

Fermilab is launching a suite of experiments in the coming years that will look for physics beyond the SM. In particular, the Muon g-2 Experiment will determine with unprecedented precision the muon anomalous magnetic moment, while the Mu2e Experiment [18] will search for neutrinoless muon to electron conversion. A combination of beam lines in the so-called Muon Campus have been designed to deliver beams sufficient for use in both experiments.

In this study, we discussed in detail the lattice design that will provide efficient capture, transport and delivery of polarized muon beams to the Muon g-2 Experiment. We have evaluated its performance by modeling the muon production target with MARS and the lattice design with G4BEAMLINE, a GEANT4 based code that fully incorporates all basic physical processes such as muon decay and spin precession. We estimate that the Muon Campus can deliver highly polarized positive muons with a rate of ∼$8 \times 10^5$ per initial 8 GeV incident proton bunch. We also found the conditions required to produce a muon beam with negligible contamination from secondary particles, where we show that three turns in the DR is sufficient for suppressing pion contamination and four turns is sufficient for separating secondary protons from the muons. Finally, both pion and muon simulated decay rates are found to have good agreement with the exponential decay law. We conclude from our numerical findings that the Muon Campus beam lines can deliver the designed beam parameters [6] for the Muon g-2 Experiment.

## ACKNOWLEDGMENTS

The authors are grateful to J. Annala, M. Berz, B. Drendel, N. Froemming, M. Korostelev, A. Liu, D. Neuffer, V. Nagaslaev and S. Werkema for many useful discussions. This work is supported by Fermi Research Alliance, LLC under Contract No. DE-AC02-07CH11359 with the United States Department of Energy.